\documentclass{ws-ijmpc}

\begin{document}

\markboth{ F. W. S. Lima}
{Majority-vote model with heterogeneous agents on square lattice}
\catchline{}{}{}{}{}

\title{Majority-vote model with heterogeneous agents on square lattice}

\author{ F. W. S. Lima}\footnote{Dietrich Stauffer 
Computational Physics Lab, Departamento de 
F\'{\i}sica,Universidade Federal do Piau\'{\i},
 Teresina, Piau\'{\i}, 64049-550, Brasil}
\address{ Dietrich Stauffer Computational Physics Lab,
 Departamento de 
F\'{\i}sica,Universidade Federal do Piau\'{\i},
Teresina,
Piau\'{\i}, 64049-550, Brasil\\
fwslima@gmail.com}

\maketitle

\begin{history}

\received{Day Month Year}

\revised{Day Month Year}

\end{history}

\begin{abstract}
We study a nonequilibrium model with up-down symmetry and
a noise parameter $q$
known as majority-vote model of M.J. Oliveira $1992$
with heterogeneous agents on square lattice.
By Monte Carlo simulations and finite-size scaling relations 
the critical exponents $\beta/\nu$, $\gamma/\nu$, and $1/\nu$
and points $q_{c}$ and
$U^*$ are obtained. After extensive simulations, we obtain
$\beta/\nu=0.35(1)$,
 $\gamma/\nu=1.23(8)$, and $1/\nu=1.05(5)$. The calculated 
values of the critical noise parameter and Binder cumulant are
 $q_{c}=0.1589(4)$ and $U^*=0.604(7)$. Within the error bars,
the exponents
obey the relation $2\beta/\nu+\gamma/\nu=2$ and  the 
results presented here demonstrate that the majority-vote model
heterogeneous agents
belongs to a different universality class than the nonequilibrium 
majority-vote 
models with homogeneous agents on square lattice.


\end{abstract}

\ccode{PACS Nos.:  05.10.Ln; 05.70.Fh; 64.60.Fr;}

\section{Introduction}

A community of people where each 
person has a characteristic (for example, 
an opinion on a particular subject and this 
opinion can be expressed in a binary form, in favor ($+1$) or against ($-1$) 
a particular issue 
in question, and this opinion can be
 influenced by the vicinity of this individual)
can be modeled using some simple 
models as the equilibrium Ising model \cite{a3,onsager} 
that has become an
excellent tool to study models of social application 
\cite{latané}. Many works these nature
are well described in a thorough review
\cite{SPSD}, a more recent summary by Stauffer \cite{stauffer1} and the 
following papers in these special issues on sociophysics in this journal. 
The majority-vote 
model (MVM) of Oliveira \cite{mario} is a nonequilibrium
model of social interaction: individuals of a certain
population make their decisions based on the opinion
of the majority of their neighbors. This model has 
been studied for several years by various 
researchers in order to model social and economic systems
\cite{zaklan,zaklan1,limanew,lima1,lima2} in regular 
structures \cite{MVM-regular,lima-malarz,santos,lima3}
 and various other complex
networks 
\cite{MVM-SW0,MVM-SW1,MVM-ERU,MVM-VD,MVM-ABD,MVM-ABU,MVM-ERD,MVM-APN}.

There are also applications to real elections in which similar models of 
opinion dynamics have been explored in the literature, 
such as Ara\'ujo {\it et al.} \cite{TVP}.

In the present work, we study the critical properties 
of MVM with random noise on a square lattice $SL$.
Here, we start with each individual or agent having 
their characteristic noise $q_{i}$ randomly selected
within a range from 0 to $q$. Thus each
agent does not have an opinion in the presence of 
a constant noise $q$ 
as in the traditional MVM
\cite{mario}, but instead each agent has 
intrinsic resistance, $q_{i}$, to the opinion of 
their neighborhood
on $SL$. The
effective dimension using the exponents ratio
$\beta/\nu$ and $\gamma/\nu$ is also determined
 for MVM with random noise. Finally, the critical exponents
calculated for this model are compared with the
results obtained by Oliveira \cite{mario}. 
\section{Model and simulation}

In the MVM on $SL$, the system dynamics traditional
is as follows. 
Initially, we assign a spin variable
$\sigma$ with values $\pm 1$ at each node of the lattice. At each step
we try to spin flip a node. The flip is  accepted with
 probability 
\begin{equation} 
w_i=\frac{1}{2}\left[ 1-(1-2q)\sigma_{i}\cdot\text{S}
\left(\sum_{j}\sigma_j\right)\right],
\label{eq_1}
\end{equation}
where $S(x)$ is the sign $\pm 1$ of $x$ if $x\neq0$, $S(x)=0$ if
$x=0$. To calculate $w_i$ our sum runs over the $k=4$ nearest
neighbors of spin $i$ on square lattice.  Eq.~(\ref{eq_1}) means that
with probability $(1-q)$ the spin will adopt the same 
state as the majority
of its neighbors. The control parameter $0\le q\le 1$
plays a role similar to the temperature in equilibrium 
systems: the smaller
$q$, the greater the probability of parallel aligning 
with the local majority.

Here, in order to make the model more realistic in a
social context we associate to each agent its 
characteristic noise {\it $q_{i}$}. Thus the agent
has not only opinion, but also an individual
resistance to the opinion of this neighborhood. Therefore,
the new rate of reversal of the spin variable is
\begin{equation} 
w_i=\frac{1}{2}\left[ 1-(1-2q_{i})\sigma_{i}\cdot\text{S}
\left(\sum_{j}\sigma_j\right)\right],
\label{eq_2}
\end{equation}
where the noise parameter {\it $q_{i}$}, associated with 
the site $i$, satisfies the probability distribution
\begin{equation} 
 P(0 < q_{i}< q) = 1/q
\label{eq_3}
\end{equation}
and takes real values randomly in the interval [0, q].

To study the critical behavior of the model we define
the variable $m\equiv\sum_{i=1}^{N}\sigma_{i}/N$ ($N=L\times L$).
In particular, we are interested in the magnetization
 $M$, susceptibility $\chi$ and the reduced fourth-order
 cumulant $U$
\begin{subequations}
\label{eq-def}
\begin{equation}
M_{L}(q)\equiv \langle|m|\rangle,
\end{equation}
\begin{equation}
\chi_{L}(q)\equiv N\left(\langle m^2\rangle-\langle m \rangle^2\right),
\end{equation}
\begin{equation}
U_{L}(q)\equiv 1-\dfrac{\langle m^{4}\rangle}{3\langle m^2 \rangle^2},
\end{equation}
\end{subequations}
where $\langle\cdots\rangle$ stands for a thermodynamic average.
The results are averaged over the $N_{\text{run}}$ 
independent simulations. 
 
These quantities are functions of the noise parameter
 $q$ and obey the finite-size scaling relations
\begin{subequations}
\label{eq-scal}
\begin{equation}
\label{eq-scal-M}
M_{L}(q)=L^{-\beta/\nu}f_m(x),
\end{equation}
\begin{equation}
\label{eq-scal-chi}
\chi_{L}(q)=L^{\gamma/\nu}f_\chi(x),
\end{equation}
\begin{equation}
\label{eq-scal-dUdq}
\frac{dU_{L}(q)}{dq}=L^{1/\nu}f_U(x),
\end{equation}
where $\nu$, $\beta$, and $\gamma$ are the usual critical 
exponents, $f_{m,\chi,U}(x)$ are the finite size scaling functions with
\begin{equation}
\label{eq-scal-x}
x=(q-q_c)L^{1/\nu}
\end{equation}
\end{subequations}
being the scaling variable. Therefore, from the size 
dependence of $M$ and $\chi$ we obtained the exponents
$\beta/\nu$ and $\gamma/\nu$, respectively. 
The maximum value of susceptibility 
also scales as $L^{\gamma/\nu}$. Moreover, the value
 of $q^*$ for which $\chi$ has a maximum is expected 
to scale with the lattice size $L$ as
\begin{equation}
\label{eq-q-max}
q^*=q_c+bL^{-1/\nu} \text{ with } b\approx 1.
\end{equation}

Therefore, the relations \eqref{eq-scal-dUdq} and
 \eqref{eq-q-max} may be used to get the exponent
$1/\nu$. We also have applied 
the calculated exponents to the 
hyperscaling hypothesis
\begin{equation}
\label{eq-q-def}
2\beta/\nu + \gamma/\nu=D_{eff}
\end{equation}
in order to get the effective dimensionality, $D_{eff}$, 
and to improve the $\beta/\nu$ and $\gamma/\nu$ 
exponents ratio for $D_{eff}=2$
on $SL$.

We performed Monte Carlo simulation on $SL$
 with various lattice 
sizes $L$ (100, 200, 300, 400, 500 and
 1000).
We took $2\times 10^5$ Monte Carlo steps (MCS) to 
make the system reach the steady state, and then 
the time averages are estimated over the next $2\times 10^5$ MCS.
One MCS is accomplished after all the $N$ spins are 
investigated whether they flip or not.

The results are averaged over 
$N_{\text{run}}$ $(100\le N_{\text{run}} \le 500)$ 
independent 
simulation runs for each lattice size and for given set 
of parameters $(q,L)$.
\section{Results and Discussion}

\begin{figure}[ph]
\centerline{\psfig{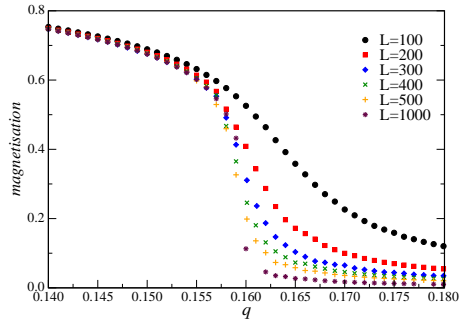}}
\vspace*{8pt}
\caption{Magnetization $M$ as a function of 
the noise parameter $q$, for $L=100$, $200$, $300$,
 $400$, $500$,  and $1000$ lattice size. \label{f1}}
\end{figure}

\begin{figure}[ph]

\centerline{\psfig{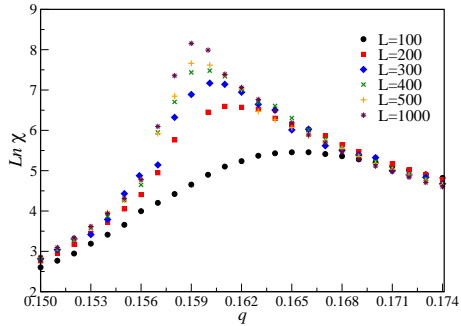}}
\vspace*{8pt}
\caption{Same as Fig.~\ref{f1}, but now for the 
susceptibility $\chi$. \label{f2}}
\end{figure}

\begin{figure}[ph]
\centerline{\psfig{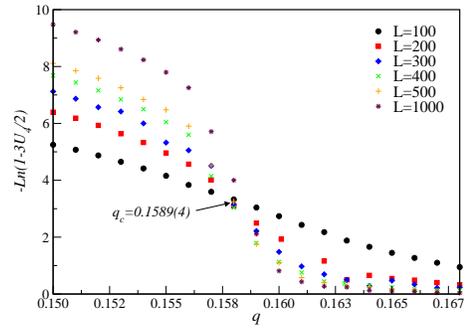}}
\vspace*{8pt}
\caption{Same as Fig.~\ref{f1}, but now for the Binder
 cumulant $U$. \label{f3}}
\end{figure}

In Figs.~\ref{f1}, ~\ref{f2}, and ~\ref{f3}  we show the 
dependence of the magnetization $M$, susceptibility $\chi$,
 and Binder cumulant $U$ on the noise parameter $q$, 
obtained from simulations on $SL$
 with $L$ ranging from $L=100$ to $1000$ lattice size
($N=10,000$ to $1,000,000$ sites). 
The shape of $M(q)$, $\chi(q)$, and
 $U(q)$ curve, for a given value of $L$, suggests the
 presence of a second-order phase transition in the 
system. The phase transition occurs at the critical
value $q_c$ of the
 noise parameter $q$.
This parameter $q_c$ is estimated as the
 point where the $U_L(q)$ curves for different lattice sizes $L$ 
intercept each other \cite{binder}.
Then, we obtain $q_{c}=0.1589(4)$ and $U^*=0.604(7)$ for
$SL$.
\bigskip

\begin{figure}[ph]
\centerline{\psfig{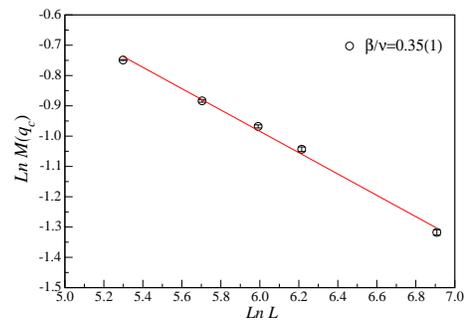}}
\vspace*{8pt}
\caption{Log-log plot of magnetization $M^*=M(q_c)$ vs. the
 linear lattice size $L$ for $SL$. \label{f4}}
\end{figure}

\begin{figure}[ph]
\centerline{\psfig{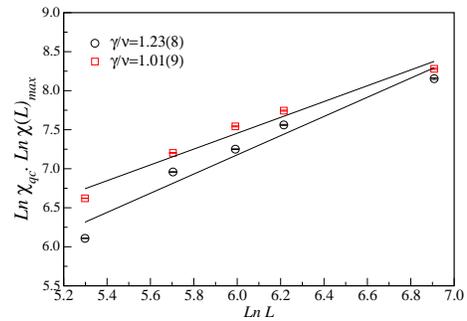}}
\vspace*{8pt}
\caption{Log-log plot of susceptibility at $q_c$ and 
$q_{\chi_{max}}(L)$ versus 
$L$ for $SL$. \label{f5}}
\end{figure}

\begin{figure}[ph]
\centerline{\psfig{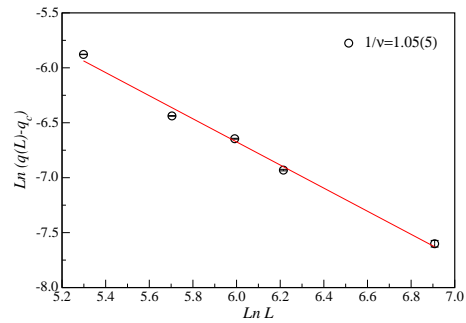}}
\vspace*{8pt}
\caption{Log-log plot of $\ln|q_c(L)-q_c|$ versus the lattice size 
$L$ for $SL$. \label{f6}}
\end{figure}
\begin{figure}[ph]

\centerline{\psfig{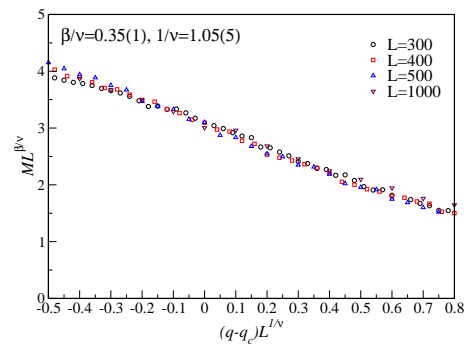}}
\vspace*{8pt}
\caption{Data collapse of the magnetisation 
{\it M} for the  lattice size $L=300$,$400$,
$500$, and $1000$ for $SL$. The exponents 
used here were
$\beta/\nu=0.35(1)$ and $1/\nu=1.05(5)$. \label{f7}}
\end{figure}

\begin{figure}[ph]
\centerline{\psfig{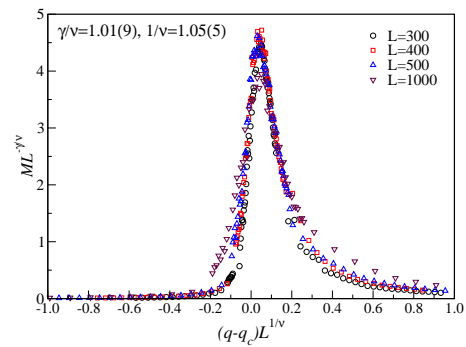}}
\vspace*{8pt}
\caption{Data collapse of the susceptibility for the  lattice 
size $L=300$,$400$,
$500$, and $1000$ for $SL$. The exponents 
used here were 
$\gamma/\nu=1.01(9)$ and $1/\nu=1.05(5)$. \label{f8}}
\end{figure}

In  Fig.~\ref{f4} we plot the dependence of the magnetization 
$M^*=M(q_c)$ vs. the lattice size $L$.
The slope of curve corresponds to the exponent ratio $\beta/\nu$
 according to Eq. (3a).
The obtained exponent is $\beta/\nu= 0.35(1)$ for our $SL$.

The exponent ratio $\gamma/\nu$ at $q_{c}$ and
$q_{\chi_{max}}(L)$ is obtained from 
the slope of the straight line with $\gamma/\nu= 1.23(8)$ and
$1.01(9)$, respectively 
as presented in Fig. \ref{f5} for $SL$. 

To obtain the critical exponent $1/\nu$, we used the scaling relation (6).
The calculated  value of the exponent $1/\nu$ are $1/\nu=1.05(5)$ 
for $SL$ (see Fig. \ref{f6}).
We plot $ML^{\beta/\nu}$ versus $(q-q_{c})L^{1/\nu}$ in Fig.
 \ref{f7} using the critical
exponents $1/\nu=1.23(8)$ and $\beta/\nu=0.35(1)$ for lattice 
size $L=300$,$400$,
$500$, and $1000$ for $SL$. The good collapse of 
the curves for five different lattice sizes corroborates the estimate 
for $q_c$ and the critical exponents $\beta/\nu$ and $1/\nu$.

In Fig. \ref{f8} we plot $\chi L^{-\gamma/\nu}$ 
versus $(q-q_{c})L^{1/\nu}$
 using the critical exponents $\gamma/\nu=1.01(9)$ and $1/\nu=1.05(5)$ 
for lattice size $L=300$, 400,
500, and 1000 for $SL$. Again, the good collapse of 
the curves for five different lattice size corroborates the extimation
 for $q_c$ and the critical exponents $\gamma/\nu$ and 
$1/\nu$.
\section{Conclusion} 

Finally, we remark that our MC results 
obtained on $SL$ for
 MVM with random noise show that critical 
exponent ratios $\beta/\nu=0.35(1)$ and
 $\gamma/\nu=1.01(9)$
are different from the results of MVM 
for regular
lattice $\beta/\nu=0.125(5)$ and 
$\gamma/\nu=1.73(5)$ \cite{mario} and 
equilibrium 2D Ising model 
\cite{onsager}. On the other hand, 
we show also that the critical 
exponent $1/\nu=1.05(5)$ and
Binder cumulant $U^*=0.604(7)$ are similar  
to the MVM for regular
lattice \cite{mario}.
We also showed that the effective dimension 
$D_{eff}$ (within error bars) is close 
to $2$.
The agreement in $D_{eff}$ and $1/\nu$ but 
not in the two exponent ratios  
$\beta/\nu$ and $\gamma/\nu$ remains to be explained.  
\bigskip

\section*{Acknowledgments}
The author thanks D. Stauffer for many suggestion and 
fruitful discussions
 during the
development this work and also for reading this paper. 
We also acknowledge the
Brazilian agency CNPQ for  its financial support. This
work also was supported the system SGI Altix 1350 in the computational park
CENAPAD.UNICAMP-USP, SP-BRAZIL.

\end{document}